\documentclass[twocolumn,english,aps,prl,groupedaddress,superscriptaddress]{revtex4-1}
\usepackage{graphicx,epsfig,units}
\usepackage{xcolor} 
\usepackage{soul} 
\usepackage{amsmath,amsfonts,mathrsfs,amsbsy,bm,babel}
\usepackage[colorlinks=true,linkcolor=blue,citecolor=blue]{hyperref}

\begin{document}

\title{Two dimensional ordering and collective magnetic excitations in the dilute ferromagnetic topological insulator (Bi$_{0.95}$Mn$_{0.05}$)$_{2}$Te$_{3}$}

\author{David~Vaknin}
\affiliation{Ames Laboratory, Ames, IA, 50011, USA}
\affiliation{Department of Physics and Astronomy, Iowa State University, Ames, IA, 50011, USA}

\author{Daniel~M.~Pajerowski}
\affiliation{Neutron Scattering Division, Oak Ridge National Laboratory, Oak Ridge, TN 37831, USA}

\author{Deborah~L.~Schlagel}
\affiliation{Ames Laboratory, Ames, IA, 50011, USA}

\author{Kevin~W.~Dennis}
\affiliation{Ames Laboratory, Ames, IA, 50011, USA}

\author{Robert~J.~McQueeney}
\affiliation{Ames Laboratory, Ames, IA, 50011, USA}
\affiliation{Department of Physics and Astronomy, Iowa State University, Ames, IA, 50011, USA}

\begin{abstract}
 Employing elastic and inelastic neutron scattering (INS) techniques, we report on the microscopic properties of the ferromagnetism in the dilute magnetic topological insulator (Bi$_{0.95}$Mn$_{0.05}$)$_{2}$Te$_{3}$. Neutron diffraction of polycrystalline samples show the ferromagnetic (FM) ordering is long-range within the basal plane, and mainly two-dimensional (2D) in character with short-range correlations between layers below $T_{\mathrm{C}} \approx 13$ K. Remarkably, we observe gapped and collective magnetic excitations in this dilute magnetic system. The excitations appear typical of quasi-2D FM systems despite the severe broadening of short wavelength magnons which is expected from the random spatial distribution of Mn atoms in the Bi planes.  Detailed analysis of the INS provide the average values for exchange couplings which are consistent with reports of carrier-mediated interactions. 
 \end{abstract}
\maketitle
The key functionality of dissipationless transport in topological insulators (TI) requires bulk ferromagnetic (FM) ordering that couples to Dirac-like topological electronic states at the surface \cite{Tokura19}.  Time-reversal-symmetry-breaking, induced by the FM ordering, gaps these surface states, giving rise to quantum anomalous Hall effect (QAHE) with chiral edge modes that can form dissipationless spin polarized currents \cite{Liu16}. Recent approaches to achieve the QAHE utilize dilute magnetic ions, such as Cr, V, or Mn, which are introduced into the tetradymite TI systems, Bi$_{2}$Te$_{3}$, Bi$_{2}$Se$_{3}$, and Sb$_{2}$Te$_{3}$ at the level of a few atomic percent and result in bulk FM order with $T_{\mathrm{C}}$ as high as 20 K \cite{Dyck02, Choi04, Dyck05, Hor10}.  Indeed, the QAHE has been observed in thin films of Bi$_{2}$Te$_{3}$ where bulk FM is induced using this approach \cite{Zhang13, Chang13, Chang15}.  However, the QAHE is observed only at $\sim$100 millikelvin, a situation which could be improved by increasing both $T_{\mathrm{C}}$ and the homogeneity of the FM state \cite{Tokura19}.

To optimize materials that exhibit these phenomena at higher temperatures, it is imperative to understand the microscopic nature of the magnetism in bulk FM-TIs.   We note that FM-TIs share common features with dilute magnetic semiconductors (DMS) for which much higher $T_{\mathrm{C}}$'s have been realized \cite{Dietl14}.  Like many DMS, such as Mn-doped GaAs, the location of solute atoms, substitutional or interstitial, and the role of solute clustering are important and sometimes raise controversial issues.  In addition, the operative microscopic mechanism of the magnetic exchange in dilute magnetic TIs, which has been described by the carrier-mediated Ruderman-Kittel-Kasuya-Yosida (RKKY) exchange  and van Vleck exchange in insulating compositions, is an open question \cite{Tokura19}.  Inelastic neutron scattering (INS) measurements can provide insight into the length- and energy-scales of the spin correlations in FM-TIs.  However, the diluteness of FM-TIs can make such measurements challenging.  For example, INS measurements performed on DMS systems have not yielded information about collective magnetism \cite{Kepa03}. 

Here, we report on INS measurements of the dilute FM-TI (Bi$_{0.95}$Mn$_{0.05}$)$_{2}$Te$_{3}$.  Despite the experimental challenges confronted in studying dilute systems with neutron scattering, we find evidence for long-range and quasi-2D FM order along with gapped, collective magnetic excitations in the FM ground state. Features of the dispersive magnons are severely broadened, likely due to random disorder.  Nonetheless, simple model calculations for a periodic triangular bilayer of exchange-coupled local-moment magnetic ions capture the essential features of the INS data, providing key magnetic energy scales.  Above $T_{\mathrm{C}}$, persistent 2D spin correlations are observed with paramagnon character.  These measurements provide unequivocal evidence of collective magnon excitations not yet observed in either DMS or dilute FM-TI systems.

(Bi$_{1-x}$Mn$_{x}$)$_{2}$Te$_{3}$ is a prototypical dilute FM-TI system.  For bulk samples with concentrations $x \leq 0.05$, Mn ions are reported to substitute randomly for Bi in the triangular layers of the tetradymite structure and FM order sets in with $T_{\mathrm{C}}$ up to 12 K \cite{Hor10, Watson13}.   The characterization of bulk samples with $x<0.05$ by magnetization, muon spin resonance, scanning probes \cite{Hor10, Watson13}, and electron spin resonance (ESR) \cite{Zimmerman17, Talanov17} are all consistent with a homogeneous FM phase.  For $x>0.05$, a solubility limit is reached whereby Mn ions cluster and form heterogeneous Mn-rich regions \cite{Hor10, Watson13}. The RKKY mechanism is suggested by $p$-type metallic conductivity \cite{Hor10} and also first-principles calculations that find a weakly bound impurity band \cite{Niu11, Li14}.  Analysis of the critical fluctuations in the paramagnetic phase with ESR data are consistent with a long-range RKKY mechanism \cite{Zimmerman17, Talanov17}.  Thus, a possible protocol to increase $T_{\mathrm{C}}$ is to increase the impurity binding energy sufficiently to induce a metal-insulator transition \cite{Chakraborty13,Bouzerar15} which could deliver bulk insulating properties and high $T_{\mathrm{C}}$, a desirable combination for QAHE \cite{Yu10}.

\begin{figure}
\includegraphics[width=0.9\linewidth]{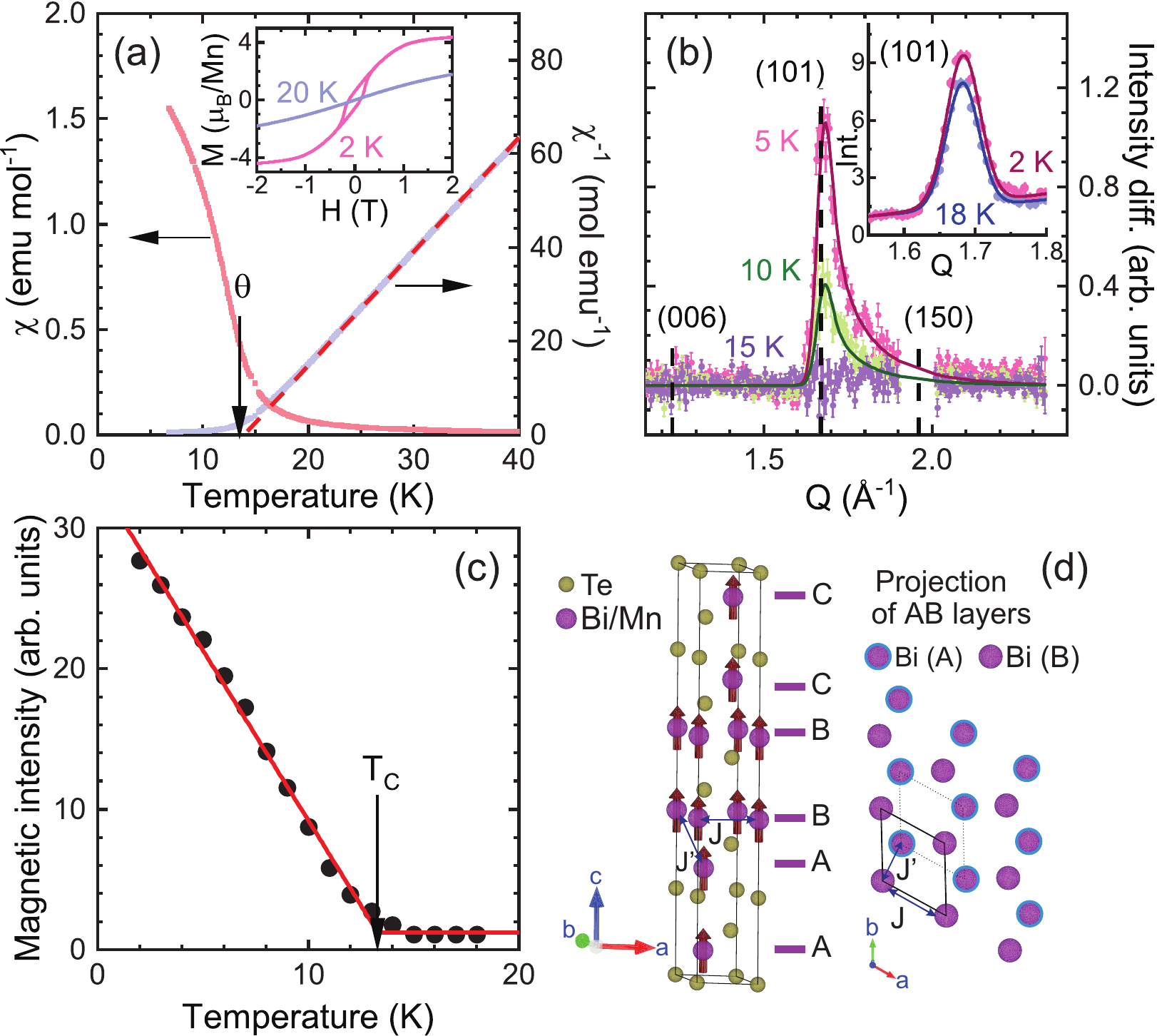}
\caption{\footnotesize (a) Zero-field-cooled magnetic susceptibility, $\chi\equiv M/H$, and 1/$\chi$ versus temperature for polycrystalline (Bi$_{0.95}$Mn$_{0.05}$)$_{2}$Te$_{3}$ at $H =0.1$ T. The red dashed line represents a Curie-Weiss fit to $1/\chi$ in between 20 and 50 K.  The inset plots $M$ vs. $H$ at 2 K and 20 K, showing hysteresis and a saturation moment of 4.4 $\mu_{B}$ per Mn at 2 K. (b) Elastic neutron scattering measured on CNCS with $E_{i}=3.32$ meV at $T =$ 5, 10, and 15 K after subtraction of the $T=18$ K signal.  Lines correspond to fits to a Warren lineshape. The inset shows the nuclear plus ferromagnetic signal at $T=2$ K superimposed on the (101) nuclear peak measured at 18 K. (c) The squared FM order parameter obtained from the  magnetic integrated intensity of the (101) peak at several temperatures.  The red line is a fit to the data assuming a mean-field-like transition.  (d) The average magnetic structure of (Bi$_{1-x}$Mn$_{x}$)$_{2}$Te$_{3}$ and key effective exchange interactions that occur between magnetic ions within the bilayers of the quintuple-layer structure.}
\label{fig1}
\end{figure}

 We prepared 25 grams each of polycrystalline Bi$_{2}$Te$_{3}$ and (Bi$_{0.95}$Mn$_{0.05}$)$_{2}$Te$_{3}$ using the approach described in the Supplemental Material (SM)  \cite{SM}.  Electron energy dispersive spectroscopy measurements are consistent with the nominal Mn composition of $x=0.05$. Curie-Weiss fits to the magnetic susceptibility measurements [Fig.~\ref{fig1}(a)] confirm FM order with a Weiss temperature of $\theta =13.5(1)$ K and an effective moment of $\mu_{eff}=4.4(1)$ $\mu_B$ per Mn, consistent with previous reports \cite{Hor10, Watson13}.  Magnetization measurements [inset of Fig.~\ref{fig1}(a)] reveal a saturation moment of 4.4 $\mu_{B}$ per Mn at $T=$ 2 K.

The elastic and inelastic neutron scattering experiments were conducted using the Cold Neutron Chopper Spectrometer (CNCS) \cite{Ehlers11,Ehlers16} at the Spallation Neutron Source at Oak Ridge National Laboratory.  For a successful INS observation of the weak magnetic signal from dilute Mn ions, it is critical to accurately subtract signals due to phonon scattering and other instrumental background contributions.  For this reason, we chose to study polycrystalline samples of the parent and doped compositions which have identical shapes that are fully illuminated by the neutron beam.  The polycrystalline samples were sealed under helium in a cylindrical aluminum cans and mounted at the cold tip of a rod that was inserted in a liquid helium dewar (an 'orange' wet $^{4}$He cryostat).  INS experiments were performed at the 2 -- 20 K temperature range. The data were collected with the 'high-flux-mode' and fixed incident neutron energies of $E_{i}=$ 1.55, 3.32, and 12.00 meV, which have approximately gaussian full-width-half-maximum instrumental elastic resolutions of 0.04, 0.11, and 0.68 meV, respectively.  

The MANTID software package \cite{Arnold14} was used to reduce the time-of-flight data sets and to produce the scattering function $S(Q,E)$, where $Q$ is the magnitude of the momentum transfer and $E$ is the energy transfer of the neutron.  A direct subtraction of the INS data measured on the parent compound from that of the Mn-doped composition, with no further corrections, provides an unambiguous signal from the magnetic fluctuations (see Fig.~S4 in the SM \cite{SM}).   For further visualization, the DAVE software package was used \cite{Azuah09}.  Miller indices ($H,K,L$) that are used throughout the manuscript to describe reciprocal space are defined with respect to the hexagonal axes.
\begin{figure*}
\includegraphics[width=1.\linewidth]{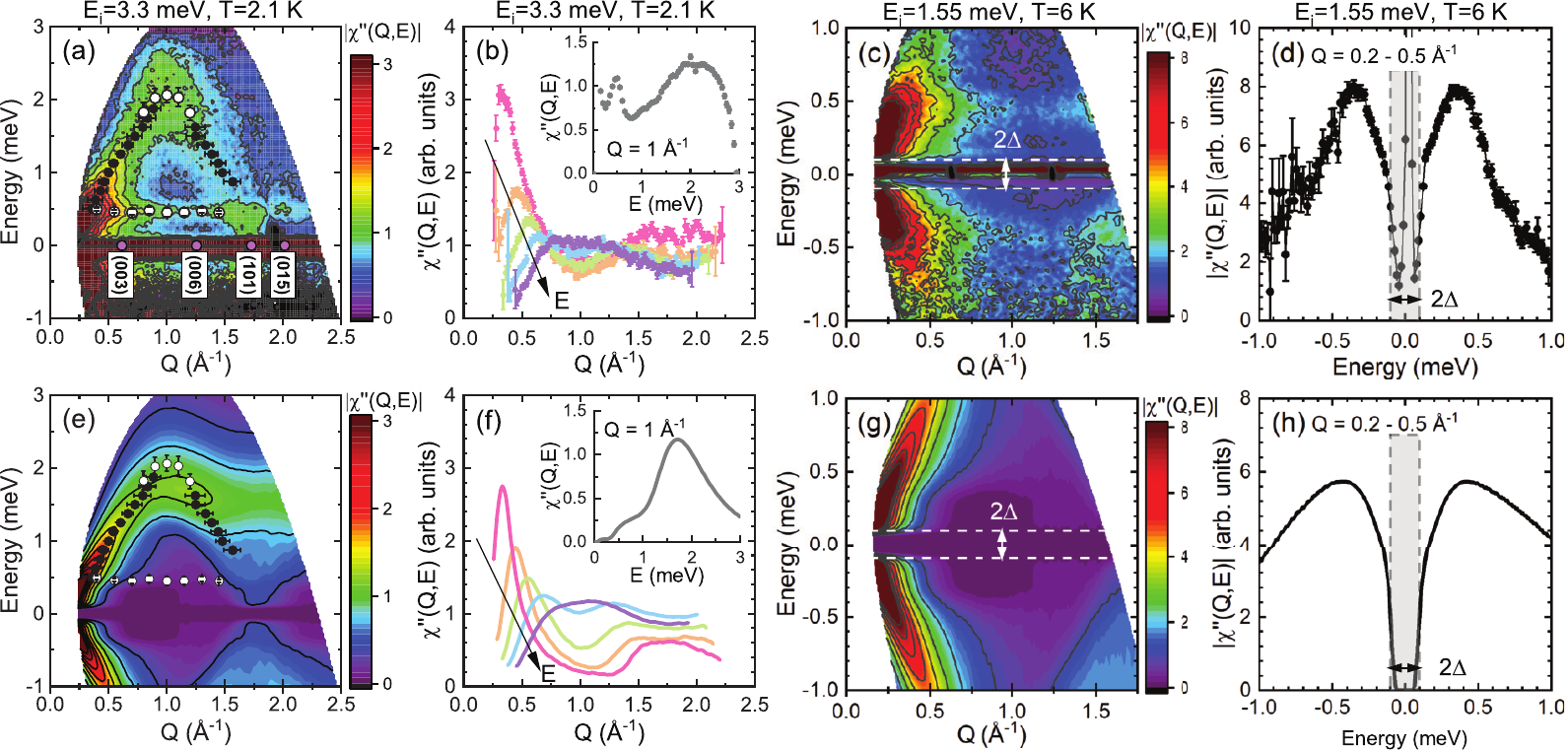}
\caption{\footnotesize (a)  INS measurements of the dynamical magnetic susceptibility, $\chi''(Q,E)$, of (Bi$_{0.95}$Mn$_{0.05}$)$_{2}$Te$_{3}$ at $T = 2.1$ K using the CNCS spectrometer. (b) A series of $Q$-cuts with $E=$ 0.6, 0.9, 1.2, 1.5 and 1.8 meV at $E_{i}=3.32$ meV that highlight the dispersive character of the magnon band.  Several $Q$-cuts were fit to a Lorentzian profile, as described in the SM \cite{SM}, and the peak values are indicated by filled circles in panel (a).  The inset in panel (b) is an $E$-cut at $Q=1$ \AA$^{-1}$ showing a  sharp, dispersionless excitation near 0.45 meV and a broad collective magnon mode at $E\approx 2$ meV.  Fits to similar $E$ cuts are indicated by empty circles in Panel (a). (c) CNCS data taken at $T=6$ K with $E_{i}=1.55$ meV and (d) $E$-cut at low-$Q$ show evidence of a spin gap with $\Delta=0.1$ meV.  (e) The dynamical magnetic susceptibility, $\chi''(Q,E)$, calculated from a damped bilayer model with $S\langle J \rangle = 0.2$ meV. $S\langle J' \rangle = 0.06$ meV, and $SD = 0.05$ meV.  The spin waves are averaged over all possible propagation directions as appropriate for a polycrystalline sample.  Kinematic restrictions appropriate for the CNCS spectrometer with $E_{i}=3.32$ meV are applied.  Filled and empty circles correspond to fits to the INS data from panel (a). (f) $Q$-cuts of the bilayer model at energy transfers of 0.6, 0.9, 1.2, 1.5 and 1.8 meV (identical cuts to those of panel (b)). The inset to panel (f) shows an energy cut from the bilayer model at $Q=1$ \AA$^{-1}$ for comparison to panel (b). (g) Low energy, low-Q dependence of the bilayer model with $E_{i}=1.55$ meV and (h) $E$-cut at low-$Q$ can be compared to the spin gap data in panels (c) and (d), respectively.
In (a), (c), (e), and (g), false color images and contour lines represent the neutron intensities as a function of $Q$ and $E$.}
\label{fig2}
\end{figure*}

 The inset of Fig.~\ref{fig1}(b) shows that the (101) diffraction peak intensity of the $x=0.05$ sample at $T=$ 2 K is enhanced relative to the 18 K peak due to FM ordering below $T_{\mathrm{C}}$.  The absence of magnetic intensity at (003) and (006) is evidence that the ordered moments have an easy-axis oriented perpendicular to the layers \cite{Hor10, Watson13}, as shown in Fig.~\ref{fig1}(d).  Figure \ref{fig1}(b) displays the (101) diffraction signal at various temperatures minus the same scan taken at 18 K. The magnetic peak near (101) has an asymmetric Warren lineshape, which is commonly found in powder diffraction from 2-D crystalline systems \cite{Warren41}.  Fits of the data to a Warren lineshape, as described in the SM \cite{SM} and shown in Fig.~\ref{fig1}(b), find a resolution limited correlation length ($> 300$~\AA) within a layer, and a very short interlayer correlation length of only $\sim7.5$~\AA~which confirms a strongly 2-D magnetic character.  The integrated intensities of the (101) magnetic peaks as a function of temperature result in a mean-field-like squared order parameter, $M^{2} \propto (T_{C}-T)$, with $T_{\mathrm{C}} =$ 13.3(2) K as shown in Fig.~\ref{fig1}(c).  The SM \cite{SM} describes details of the analysis of the magnetic diffraction and structural refinements of the elastic diffraction data, where we find that the extracted average ordered magnetic moment is larger than that expected for a Mn$^{2+}$ moment. 

INS captures the momentum ($\hbar Q$) and energy (E) dependencies of the magnetic excitations through the cross-section for magnetic scattering, $S(Q,E)$.  $S(Q,E)$ is obtained for (Bi$_{0.95}$Mn$_{0.05}$)$_{2}$Te$_{3}$ after subtracting the non-magnetic Bi$_{2}$Te$_{3}$ data (see SM \cite{SM}) and is proportional to the imaginary part of the dynamical magnetic susceptibility, $\chi''(Q,E)$, 
\begin{equation}
S(Q,E) \propto f^{2}(Q) e^{-2W}[1+n(E)] \chi''(Q,E).
\label{magcross}
\end{equation}
Here $n(E)$ is the Bose occupancy factor and $f(Q)$ is the magnetic form factor for Mn$^{2+}$. The dynamical susceptibility is extracted in arbitrary units after dividing out the Bose and magnetic form factors.  We assume that the Debye-Waller factor ($e^{-2W}$) is one.

Figure~\ref{fig2}(a) shows the dynamical susceptibility for (Bi$_{0.95}$Mn$_{0.05}$)$_{2}$Te$_{3}$ measured on CNCS with $E_{i}=3.32$ meV and $T=2.1$ K.  A broad, dispersive magnon branch emanates from $Q=0$, reaches a maximum of $E\approx$ 2 meV near the zone boundary of the triangular layer [$Q$(0.5,0,0) = 0.84 \AA$^{-1}$], and returns to $E\approx 0$ near the $\Gamma$-point [$Q$(101) = 1.7 \AA$^{-1}$].  The insensitivity of the dispersion to Brillouin zone centers and boundaries along (00$L$) highlights the quasi-2D character of the magnetism. A low-energy dispersionless mode is also seen at $E \approx 0.5$ meV in Fig.~\ref{fig2}(a).  

Data points overlayed on Fig.~\ref{fig2}(a) are obtained from fits to a series of constant-$Q$ and constant-$E$ cuts similar to those shown in Fig.~\ref{fig2}(b), as described in the SM \cite{SM}.  These fits find a maximum magnon energy of $\hbar\omega_{max}\approx 2$ meV and a dispersionless branch at $\hbar\omega_{0}\approx$ 0.45 meV.  As discussed below, the dispersionless branch can be explained either as an optical magnon branch or as a localized magnetic excitation from isolated Mn-Mn pairs.  Figures~\ref{fig2}(c) and  \ref{fig2}(d) focus on lower energy excitations using $E_{i}=1.55$ meV and $T=6$ K.  Both the image plot and the low-$Q$ $E$-cut identify a spin gap with $\Delta = 0.1$ meV. Fig.~S5 \cite{SM} indicates that the dynamical susceptibility at $T=$ 2.1 K and 6 K are nearly identical.

The INS data suggest a minimal model for the spin dynamics.  We assume that Mn ions occupy Bi sites and are magnetically coupled within a bilayer of the quintuple-layer structure.  The lack of dispersive features near the (00L) Bragg peaks allows us to assume that negligible coupling occurs between the bilayers [see Fig.~\ref{fig1}(d)].  Despite the dilute and disordered nature of the system, a periodic Heisenberg model with nearest-neighbor (NN) coupling captures many of the essential features of the data, as follows,  
\begin{equation}
\begin{split}
 H=-\langle J \rangle\sum_{\langle ij \rangle, A}\textbf{S}_{Ai} \cdot \textbf{S}_{Aj} - \langle J \rangle\sum_{\langle ij \rangle, B}\textbf{S}_{Bi} \cdot \textbf{S}_{Bj} \\
 - \langle J' \rangle\sum_{\langle ij \rangle, AB}\textbf{S}_{Ai} \cdot \textbf{S}_{Bj}-D\sum_i(S_i^z )^2.     
\label{hamiltonian}
 \end{split}
\end{equation}
Here, $A$ and $B$ label the two layers within a bilayer that contain identical spins of magnitude $S$, $\langle J \rangle$ is the average NN FM exchange within a single $A$ or $B$ layer, $\langle J' \rangle$ is the average NN FM exchange between $A$ and $B$ layers, and $D$ is the single-ion anisotropy. 

Within linear spin wave theory of the Heisenberg model (see SM \cite{SM}), the key energy scales are $\hbar\omega_{max}=9S\langle J \rangle+3S\langle J' \rangle\approx 2$ meV, $\hbar\omega_{0}=\Delta+6S\langle J' \rangle\approx 0.45$ meV, and $\Delta = 2SD \approx 0.1$ meV. From these relations, we estimate that $S\langle J \rangle=0.20$ meV, $S\langle J' \rangle=0.06$ meV, and $SD=0.05$ meV.  Using these values and assuming $S \approx 2$, the mean-field Curie temperature, $T_{\mathrm{C}}=S(S+1)(6\langle J \rangle+3\langle J' \rangle)/3k_{\mathrm{B}}\approx$ 16 K, is in line with the measured transition temperature.  These energy scales are also consistent with estimates based on ESR data, although ESR measurements could not be performed in the ordered state due to the spin gap \cite{Zimmerman17, Talanov17}.

Using these exchange values, the dispersion and the polycrystalline-averaged neutron scattering intensities are calculated using Monte Carlo integration methods \cite{McQueeney08}.  While the general features of the INS data are captured with this model, the calculated intensities are generally too sharp, since dilution and disorder are not accounted for in our model (see Fig.~S8 \cite{SM}).  One expected consequence of disorder is the damping of magnons with wavelengths that are shorter than the mean spacing between moments. The convolution of these calculations with a phenomenological energy-dependent damping parameter ($\gamma(E) = \sqrt{\sigma_{res}^{2}+(\beta E)^{2}}$ where $\sigma_{res}=$ 0.045 meV is the instrumental energy resolution and $\beta=$ 0.28) improves the agreement.   

As shown in Fig.~\ref{fig2}(e)--(h), many qualitative features of the data are captured by the damped bilayer model.  For example, comparison of Figs.~\ref{fig2}(b) and \ref{fig2}(f) shows that the oscillation and width of constant-$E$ $Q$-cuts are similar, but the energy dependence of the magnetic spectral weight is not. Disagreement between the data and model for $Q>1.7$ \AA$^{-1}$ likely arise from the subtraction of intense phonon contributions (see Fig.~S4 \cite{SM}).  Finally, in the bilayer model, the localized mode at $\hbar\omega_{0}$ is interpreted as an optical magnon branch caused by opposite spin precession between the $A$ and $B$ layers.  However, the localized mode is much sharper and more intense in the data which could indicate that it may alternatively be due to local spin dimer excitations from isolated Mn-Mn pairs.  Similar spin dimer excitations have been observed in INS studies of DMS systems \cite{Kepa03}.  

\begin{figure}
\includegraphics[width=1.\linewidth]{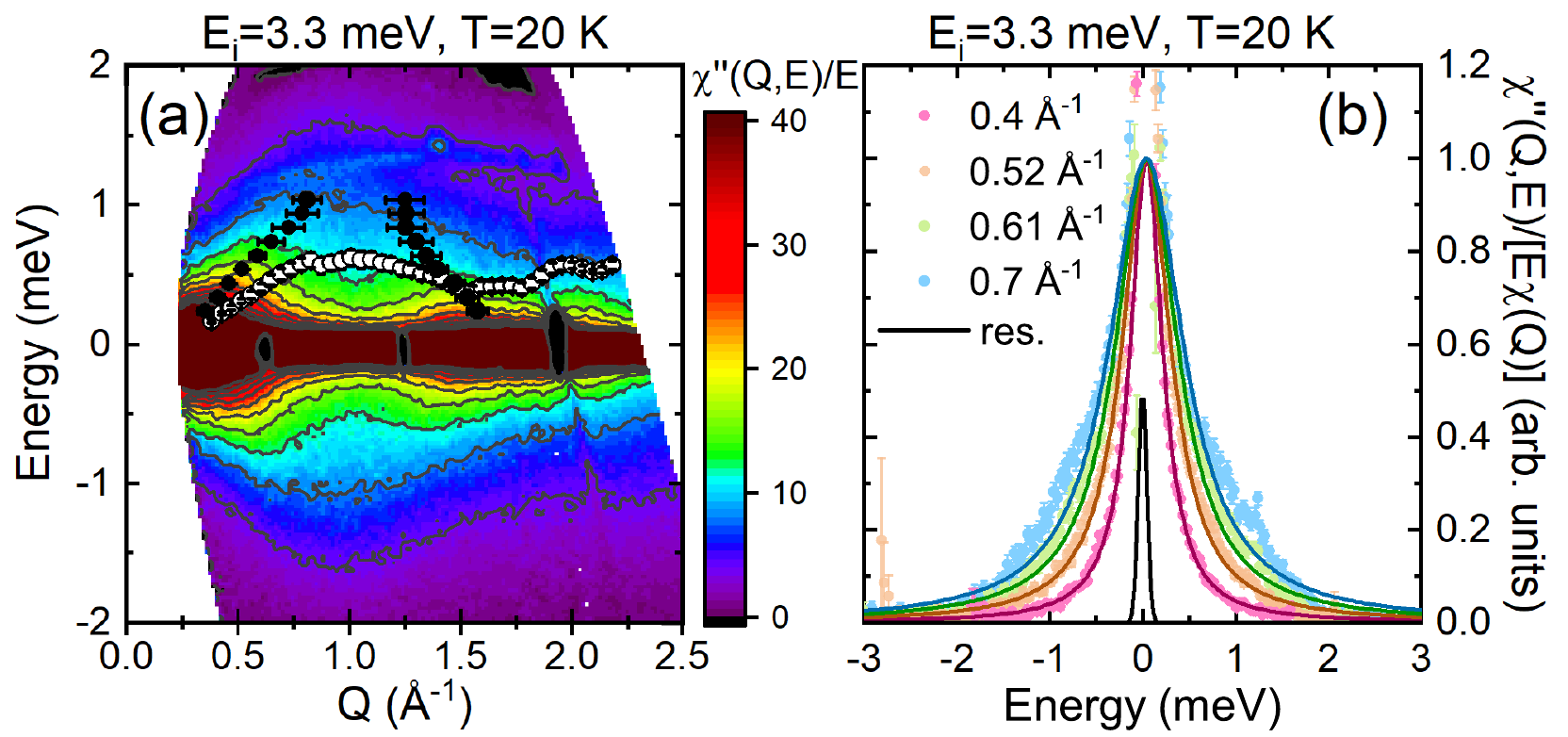}
\caption{\footnotesize (a) The power spectrum [$\chi''(Q,E)/E$] of (Bi$_{0.95}$Mn$_{0.05}$)$_{2}$Te$_{3}$ above $T_{\mathrm{C}}$ at $T = 20$ K with $E_{i}=3.32$ meV. (b) A series of $E$-cuts through the power spectrum at  $Q=$ 0.4, 0.52, 0.61 and 0.7 \AA $^{-1}$ along with Lorentzian fits.  The fitting results find a $Q$-dependent relaxation energy, $\Gamma(Q)$, shown as empty circles in panel (a) consistent with FM fluctuations. Panel (a) also shows the paramagnon dispersion (fitted maxima in the power spectrum) as filled circles.  See SM for more details \cite{SM}.}
\label{fig4}
\end{figure}

The spin fluctuations above $T_{\mathrm{C}}$ are characterized by relaxational dynamics typical of a nearly ordered FM system.  Figure \ref{fig4}(a) shows the power spectrum [$\chi''(Q,E)/E$] of the excitations at $T=20$ K and Fig.~\ref{fig4}(b) shows a gapless response where cuts at different $Q$-values are characterized by a relaxational (Lorentzian) form
\begin{equation}
 \frac{\chi''(Q,E)}{E}=\frac{\chi(Q)\Gamma(Q)}{E^{2}+\Gamma(Q)^{2}}.
\end{equation}
Here, $\Gamma(Q)$ is the relaxation rate and $\chi(Q)$ is the static susceptibility.  Fitted values of $\Gamma(Q)$, indicated by the open circles in Fig.\ \ref{fig4}(a), show a strong $Q$-dependence as expected for FM fluctuations (see SM \cite{SM}) \cite{Moriya}. The combined $Q$-dependences of $\chi(Q)$ and $\Gamma(Q)$ result in maxima in the paramagnetic response (full circles in Fig.\ \ref{fig4}(a)) commonly referred to as paramagnons.  These features highlight that, both above and below $T_{\mathrm{C}}$, qualitative features of the magnetic excitations in (Bi$_{0.95}$Mn$_{0.05}$)$_{2}$Te$_{3}$ are typical of a FM system despite their dilute and disordered nature.

Our results serve as tests of \textit{ab-initio} theoretical calculations that provide a basis for estimating the strength of magnetic interactions in various dilute FM-TIs \cite{Henk12, Zhang13_2, Verginory14, Li14}.  With regard to Mn dopants, all of these studies support hole-mediated exchange with predominantly FM interactions at short distances within a quintuple layer, consistent with our bilayer model analysis. More detailed first-principles estimates of the pairwise exchange interactions find both $J$ and $J'$ to be in the range from 2--4 meV with $J/J' \approx 1.5$ \cite{Henk12}.  This is in reasonable agreement with our bilayer model where (assuming $S\simeq 2$) we estimate that $J \approx S\langle J \rangle/xS = 2$ meV and $J' \approx S\langle J' \rangle/xS = 0.6$ meV, although the experimental value for $J'$ is subject to our interpretation that the dispersionless mode at 0.45 meV is an optical magnon branch.   The prediction of a substantial FM coupling between bilayers Ref.~\cite{Verginory14} is not corroborated here.  Our discovery that magnetism in (Bi$_{0.95}$Mn$_{0.05}$)$_{2}$Te$_{3}$ is strongly 2D is likely to carry over to other magnetic tetradymite systems and presents a hard limit for the highest achievable $T_{\mathrm{C}}$.

Finally, we mention recent reports of segregation and intercalation of Mn into 2D sheets in thin film samples of Bi$_{2}$Te$_{3}$ and Bi$_{2}$Te$_{3}$ with dilute Mn substitution \cite{Rienks18}, thereby forming septuple layers akin to MnBi$_{2}$Te$_{4}$  \cite{Otrokov18}.  Although there is as yet no experimental evidence that similar segregation occurs in bulk samples, including our own, we note that our INS data could be interpreted in such a scenario.  In this scenario, $S\langle J \rangle$ would more closely represent the in-plane NN FM coupling expected for MnBi$_{2}$Te$_{4}$.

In summary, the microscopic nature of the magnetism in the dilute FM-TI (Bi$_{0.95}$Mn$_{0.05}$)$_{2}$Te$_{3}$ has been revealed by INS.  The FM order is quasi-2D, long-range within the basal plane with short-range correlations between layers.  The excitations are typical of a quasi-2D FM system, despite the dilute concentration of Mn ions. Below $T_{\mathrm{C}}$, collective magnons are severely broadened at short wavelengths.  We note that INS studies of DMS systems have shown localized excitations from Mn-Mn pairs, but have not yet revealed collective magnon modes \cite{Kepa03}.  Thus, modern INS instrumentation promises to deliver crucial information for (Bi$_{1-x}$Mn$_{x}$)$_{2}$Te$_{3}$ and related dilute FM systems. In particular, methods developed in the DMS community to analyze magnetic excitations in dilute systems \cite{Chakraborty13, Bouzerar15} should be applied to magnetic TIs and may serve to guide protocols to increase $T_{\mathrm{C}}$.

\section{Acknowledgments}
This research was supported by the U.S. Department of Energy, Office of Basic Energy Sciences, Division of Materials Sciences and Engineering. Ames Laboratory is operated for the U.S. Department of Energy by Iowa State University under Contract No. DE-AC02-07CH11358.  A portion of this research used resources at the Spallation Neutron Source, a DOE Office of Science User Facility operated by the Oak Ridge National Laboratory.
 This manuscript has been authored in part by UT-Battelle, LLC, under contract DE-AC05-00OR22725 with the US Department of Energy (DOE). The US government retains and the publisher, by accepting the article for publication, acknowledges that the US government retains a nonexclusive, paid-up, irrevocable, worldwide license to publish or reproduce the published form of this manuscript, or allow others to do so, for US government purposes. DOE will provide public access to these results of federally sponsored research in accordance with the DOE Public Access Plan (\url{http://energy.gov/downloads/doe-public-access-plan}).




\begin{thebibliography}{30}
\bibitem{Tokura19} Y. Tokura, K. Yasuda, and A. Tsukazaki, Nat. Rev. Phys. \textbf{1}, 126 (2019).
\bibitem{Liu16} C. X. Liu, S. C. Zhang, and X. L. Qi,  in Annual Review of Condensed Matter Physics, Vol. 7 (eds. M. C. Marchetti and S. Sachdev)  301 (2016).
\bibitem{Dyck02} J. S. Dyck, P. H\'{a}jek, P. Lo\v{s}t'\'{a}k, and C. Uher, Phys. Rev. B \textbf{65}, 115212 (2002).
\bibitem{Choi04} J. Choi, S. Choi, J. Choi, Y. Park, H.-M. Park, H.-W. Lee, B.-C. Woo, and S. Cho, Phys. Status Solidi (b) \textbf{241}, 1541 (2004).
\bibitem{Dyck05} J. S. Dyck, \v{C}. Dra\v{s}ar,  P. Lo\v{s}t'\'{a}k,  and C. Uher, Phys. Rev. B \textbf{71}, 115214 (2005).
\bibitem{Hor10} Y. S. Hor, P. Roushan, H. Beidenkopf, J. Seo, D. Qu, J. G. Checkelsky, L. A. Wray, D. Hsieh, Y. Xia, S. Y. Xu \textit {et al.},  Phys. Rev. B \textbf{81}, 195203 (2010).
\bibitem{Zhang13} J. Zhang, C.-Z. Chang, P. Tang, Z. Zhang, X. Feng, K. Li, L.-l. Wang, X. Chen, C. Liu, W. Duan \textit{et al.}, {Science} \textbf{339}, 1582 (2013).
\bibitem{Chang13} C.-Z. Chang, J. Zhang, X. Feng, J. Shen, Z. Zhang, M. Guo, K. Li, Y. Ou, P. Wei, L.-L. Wang \textit{et al.}, Science \textbf{340}, 167 (2013).
\bibitem{Chang15} C.-Z. Chang, W. Zhao, D. Y. Kim, H. Zhang, B. A. Assaf, D. Heiman, S.-C. Zhang, C. Liu, M. H. W. Chan, and J. S. Moodera, Nat. Mater. \textbf{14}, 473 (2015).
\bibitem{Dietl14} T. Dietl and H. Ohno,  Rev. Mod. Phys. \textbf{86}, 187 (2014).
\bibitem{Kepa03} H. K\c{e}pa, V. K. Le, C. M. Brown, M. Sawicki, J. K. Furdyna, T. M. Giebultowicz, and T. Dietl, Phys. Rev. Lett. \textbf{91}, 087205 (2003).
\bibitem{Watson13} M. D. Watson, L. J. Collins-McIntyre, L. R. Shelford, A. I. Coldea, D. Prabhakaran, S. C. Speller, T. Mousavi, C. R. M. Grovenor, Z. Salman, S. R. Giblin, \textit {et al.}, New J. Phys. \textbf{15}, 103016 (2013).
\bibitem{Zimmerman17} S. Zimmermann, F. Steckel, C. Hess, H. W. Ji, Y. S. Hor, R. J. Cava, B. Bchner, and V. Kataev, Phys. Rev. B \textbf{94}, 125205 (2016).
\bibitem{Talanov17} Y. Talanov, V.  Sakhin, E. Kukovitskii, N. Garif'yanov, and G. Teitel'baum, Appl. Magn. Reson. \textbf{48}, 143-154 (2017).
\bibitem{Niu11} C. Niu, Y. Dai, M. Guo, W. Wei, Y. Ma, and B. Huang, Appl. Phys. Lett. \textbf{98}, 252502 (2011).
\bibitem{Li14} Y. Li, X.  Zou, J. Li, J. and G. Zhou, J. Chem. Phys. \textbf{140}, 124704 (2014).
\bibitem{Chakraborty13} C. Akash, and B. Georges, Europhys. Lett. \textbf{104}, 57010 (2013).
\bibitem{Bouzerar15} G. Bouzerar, and R. Bouzerar, Comptes Rendus Physique \textbf{16}, 731-738, (2015).
\bibitem{Yu10} R. Yu, W. Zhang, H.-J. Zhang, S.-C. Zhang, X. Dai, and Z. Fang, Science \textbf{329}, 61 (2010).
\bibitem{SM} See Supplemental Material at [URL will be inserted by publisher] for details of the analysis of neutron scattering data and modeling of the magnetic excitations.
\bibitem{Ehlers11} G. Ehlers, A. Podlesnyak, J. L. Niedziela, E. B. Iverson, and P. E. Sokol, Rev. Sci. Instrum. 82, 085108 (2011).
\bibitem{Ehlers16} G. Ehlers, A. Podlesnyak, and A. I. Kolesnikov, Rev. Sci. Instrum. \textbf{87}, 093902 (2016).
\bibitem{Arnold14} O. Arnold, J. C. Bilheux, J. M. Borreguero, A. Buts, S. I. Campbell, L. Chapon, M. Doucet, N. Draper, R. Ferraz Leal, M. A. Gigg \textit{et al.}, Nucl. Instrum. Methods Phys. Res. Sect. A \textbf{764}, 156 (2014).
\bibitem{Azuah09} R.T. Azuah, L.R. Kneller, Y. Qiu, P.L.W. Tregenna-Piggott, C.M. Brown, J.R.D. Copley, and R.M. Dimeo, J. Res. Natl. Inst. Stan. Technol. \textbf{114}, 341 (2009).
\bibitem{Warren41}B. E. Warren, Phys. Rev. \textbf{59}, 693 (1941).
\bibitem{McQueeney08} R. J. McQueeney,  J. Q. Yan, S. Chang, and J. Ma, Phys. Rev. B \textbf{78}, 184417 (2008).
\bibitem{Moriya} T. Moriya, \textit{Spin Fluctuations in Itinerant Electron Magnetism.}  (Springer Berlin Heidelberg, 2012).
\bibitem{Henk12} J. Henk, M. Flieger, I. V. Maznichenko, I. Mertig, A. Ernst, S. V. Eremeev, and E. V. Chulkov, Phys. Rev. Lett. \textbf{109}, 076801 (2012).
\bibitem{Zhang13_2} J. M. Zhang, W. M. Ming, Z. G. Huang, G. B. Liu, X. F. Kou, Y. B. Fan, K. L. Wang, and Y. G. Yao, Phys. Rev. B \textbf{88}, 235131 (2013).
\bibitem{Verginory14} M. G. Vergniory, M. M. Otrokov, D. Thonig, M. Hoffmann, I. V. Maznichenko, M. Geilhufe, X. Zubizarreta, S. Ostanin, A. Marmodoro, J. Henk \textit{et al.}, Phys. Rev. B \textbf{89}, 165202 (2014).
\bibitem{Rienks18} E. D. L. Rienks, S. Wimmer, P. S. Mandal, O. Caha, J. R\r{u}\v{z}i\v{c}ka, A. Ney, H. Steiner, V. V. Volobuev, H. Groiss, M. Albu \textit{et al.}, arXiv:1810.06238  (2018).
\bibitem{Otrokov18} M. M. Otrokov, I. I. Klimovskikh, H. Bentmann, A. Zeugner, Z. S. Aliev, S. Gass, A. U. B. Wolter, A. V. Koroleva, D. Estyunin, A. M. Shikin \textit{et al.}, arXiv:1809.07389 (2018).
\end{thebibliography}
\end{document}